\documentclass[aps,prl,showpacs,superscriptaddress,floatfix,twocolumn]{revtex4-1}
\usepackage{times}
\usepackage{graphicx}
\usepackage{amsmath}
\usepackage{amssymb}
\usepackage{subfigure}
\usepackage{color}
\usepackage{epstopdf}

\begin{document}

\title{Induced magnetization in La$_{0.7}$Sr$_{0.3}$MnO$_3$/BiFeO$_3$ superlattices}

\author{S. Singh}
\affiliation{Los Alamos National Laboratory,
Los Alamos, New Mexico 87545, USA}
\affiliation{Solid State Physics Division, Bhabha Atomic Research Center, Mumbai 400085, India}

\author{J. T. Haraldsen}
\affiliation{Los Alamos National Laboratory,
Los Alamos, New Mexico 87545, USA}
\affiliation{Department of Physics and Astronomy, James Madison University, Harrisonburg, Virginia 22807, USA}

\author{J. Xiong}
\affiliation{Los Alamos National Laboratory,
Los Alamos, New Mexico 87545, USA}
\affiliation{State Key Lab of Electronic Thin Films and Integrated Devices, University of Electronic Science and Technology of China, Chengdu 610051, China} 

\author{E. M. Choi}
\affiliation{Department of Materials Science and Metallurgy, University of Cambridge, Cambridge CB2 3QZ, UK}

\author{P. Lu}
\affiliation{Sandia National Laboratory, Albuquerque, New Mexico 87185, USA}

\author{ D. Yi}
\affiliation{Department of Materials Science and Engineering, University of California at Berkeley, Berkeley, California 94720, USA }

\author{X.-D. Wen}
\affiliation{Los Alamos National Laboratory,
Los Alamos, New Mexico 87545, USA}

\author{J. Liu}
\affiliation{Department of Materials Science and Engineering, University of California at Berkeley, Berkeley, California 94720, USA }

\author{H. Wang}
\affiliation{Department of Electrical and Computer Engineering, Texas A\&M University, College Station, Texas 77843, USA}

\author{Z. Bi}
\affiliation{Los Alamos National Laboratory,
Los Alamos, New Mexico 87545, USA}

\author{P. Yu}
\affiliation{Department of Materials Science and Engineering, University of California at Berkeley, Berkeley, California 94720, USA }

\author{M. R. Fitzsimmons}
\affiliation{Los Alamos National Laboratory,
Los Alamos, New Mexico 87545, USA}

\author{J. L. MacManus-Driscoll}
\affiliation{Department of Materials Science and Metallurgy, University of Cambridge, Cambridge CB2 3QZ, UK}

\author{R. Ramesh}
\affiliation{Department of Materials Science and Engineering, University of California at Berkeley, Berkeley, California 94720, USA }

\author{A. V. Balatsky}
\affiliation{Los Alamos National Laboratory,
Los Alamos, New Mexico 87545, USA}
%\affiliation{Center for Integrated Nanotechnologies, Los Alamos National Laboratory, Los Alamos, New Mexico 87545, USA}

\author{Jian-Xin Zhu}
\affiliation{Los Alamos National Laboratory,
Los Alamos, New Mexico 87545, USA}

\author{Q. X. Jia}
\email[To whom correspondence should be addressed. \\ Electronic address: ]{qxjia@lanl.gov}
\affiliation{Los Alamos National Laboratory,
Los Alamos, New Mexico 87545, USA}

\begin{abstract}
Using polarized neutron reflectometry (PNR), we observe an induced magnetization of 75$\pm$ 25 kA/m at 10 K in a La$_{0.7}$Sr$_{0.3}$MnO$_3$ (LSMO)/BiFeO$_3$ superlattice extending from the interface through several atomic layers of the BiFeO$_3$ (BFO). The induced magnetization in BFO is explained by density functional theory, where the size of bandgap of BFO plays an important role. Considering a classical exchange field between the LSMO and BFO layers, we further show that magnetization is expected to extend throughout the BFO, which provides a theoretical explanation for the results of the neutron scattering experiment.  
\end{abstract}
\maketitle

{\it Introduction.~}  Hybrid complex oxide nanostructures with controlled geometry and dimensionality provide an unprecedented platform to introduce and explore competing physical phenomena in functional materials. In particular, artificially engineered epitaxial heterostructures
%, in which different strongly correlated materials are interfaced in layered structures, 
enable new functionalities that cannot be realized with the individual constituents. Since functionality is derived from the interactions across interfaces~\cite{NNakagawa06}, an understanding of the interfacial structure and properties is critically important to achieve the goal of prediction and control of the properties. 
Experimental results~\cite{HZheng04,AOhtomo04,ABrinkman07,NReyren07} have shown that either structural or electrostatic boundary conditions can be dominant factors in controlling the atomic, electronic, and magnetic structures of interfaces in solid-solid. The availability of various heterostructures formed by different correlated electron materials offers new opportunities for studying competing interactions between different properties (charge-ordered, ferroelectric (FE), ferromagnetic (FM), and superconducting states) at interfaces. 
%These interactions produce a delicate balance between states with very different properties, such as charge-ordered, ferroelectric (FE), ferromagnetic (FM), and superconducting states, thereby giving rise to new physical effects. 
%Noteworthy examples include an enhancement of magnetoelectric (ME) effects in some piezoelectric/ferroelectric systems by several orders of magnitude, such as BaTiO$_3$ interfaced with magnetic CoFe$_2$O$_4$~\cite{HZheng04}, and 
Noteworthy emergent behaviors at the interface between otherwise strongly insulating materials, like LaAlO$_3$ and SrTiO$_3$, which arises from possible polar discontinuities at the interfaces~\cite{AOhtomo04,ABrinkman07,NReyren07,CCen09,HYHwang12} and/or chemical intermixing across the interfaces~\cite{SAChambers06,LQiao11,SAChambers10}. The formation of an enhanced canted magnetic state in the antiferromagnet (AFM) BiFeO$_3$ (BFO) at the interface with ferromagnetic La$_{0.7}$Sr$_{0.3}$MnO$_{3}$ (LSMO) is another intriguing observation~\cite{PYu10,SMWu10}. Even though x-ray magnetic circular dichroism (XMCD) measurements have demonstrated exchange coupling between the ferromagnetic LSMO and antiferromagnetic BFO mediated through an enhanced magnetic state localized at the interface~\cite{PYu10}, quantitative measurements to test theoretical models are lacking. In this Letter, we report such measurements of the interface magnetism and explain its origin using density functional theory together with an effective exchange field model. 

\begin{figure}
\centering\includegraphics[%scale=0.28
width=1.0\linewidth,clip]{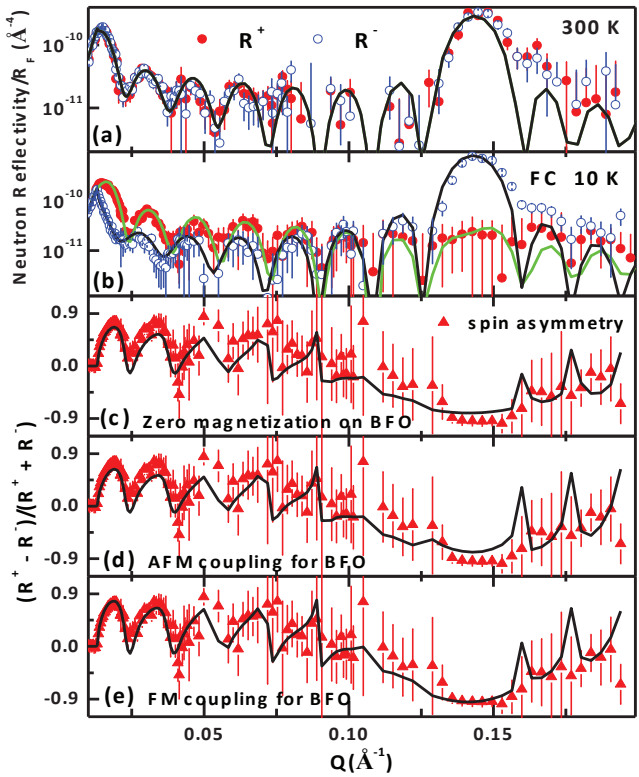}
\caption{(Color online) 
Polarized neutron reflectivity for a superlattice [(LSMO)$_6$/(BFO)$_5$]$_8$ on (100) SrTiO$_3$ substrate.  Measured data at 300 K (a) and at 10 K (b), as well as the fitting of data at 10 K with  non-magnetization (c), antiferromagnetic (d), and ferromagnetic (e) models. For the field cooled measurement, the sample was initially cooled in a field of 1 kOe from 300 K, and then measured while warming the sample at the same field. Closed and open circles are experimental data for neutron with spin parallel ($R^+$) and antiparallel ($R^-$) to the magnetic field. Both solid-black and -green lines are the best fit to the experimental data.
}
\label{fig:PNR}
\end{figure}

{\em Experiments.~} %We start with 
The synthesis of superlattices [(LSMO)$_n$/(BFO)$_m$]$_N$ on (001) SrTiO$_3$ (STO) substrates was done by pulsed laser (KrF) deposition, where the numbers $(n, m)$ of unit cell (u.c.) for LSMO and BFO as well as the stacking periodicity of $N$ were adjusted to maintain the total layer thickness in the range of 300 -- 400 \AA. Full growth details are given in the Supplemental Material (SM)~\cite{SupplMater}. Evidence for chemically and structurally well-defined interfaces over lateral dimensions of tens of nm was obtained using x-ray diffraction (Figs. S1 and S2 in Ref.~\onlinecite{SupplMater}) and high angle annular dark field (HAADF) Z-contrast microscopy (Fig. S3 in Ref.~\onlinecite{SupplMater} ). X-ray diffraction rocking curves of the superlattice (002) peak exhibit a full-width at half-maximum of 0.028$^\circ$ as compared to a value of 0.017$^\circ$ for the (002) of single crystal STO substrate. These results are evidence for a high degree of perfection of crystal structure along the growth direction. Channeling of He ions measured with Rutherford Backscattering Spectroscopy shows a value of minimum yield 3.40\% for La in the superlattice, in comparison with a value of 4.85\% for Sr in the STO substrate. This result provides additional evidence of a high degree of registry (epitaxy) between the crystal structure of the film and that of the single crystal substrate.

\begin{figure}[t]
\centering\includegraphics[%scale=0.28
width=1.0\linewidth,clip]{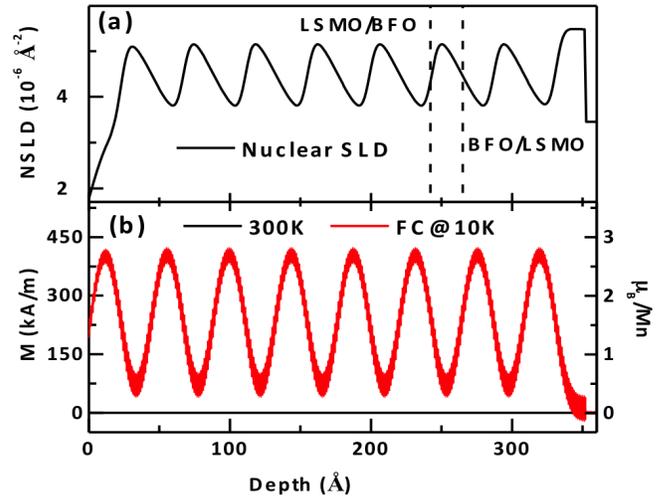}
\caption{(Color online) 
The depth profile of the characteristic parameters for a superlattice [(LSMO)$_6$/(BFO)$_5$]$_8$ on (001) SrTiO$_3$ substrate. (a) The nuclear scattering length density; and (b) The magnetization at 300 K and 10 K, which gave the best fit to
PNR data shown in Fig.~\ref{fig:PNR}. The dashed vertical line in (a) is used to distinguish the two interfaces of LSMO on BFO and BFO on LSMO.
}
\label{fig:depth_profile}
\end{figure}

In order to probe the depth-dependent structure and magnetization of the superlattice, we carried out specular x-ray reflectivity (XRR, Fig. S2 in Ref.~\onlinecite{SupplMater}) and polarized neutron reflectivity (PNR) measurements. XRR and PNR are two non-destructive techniques that provide quantitative measures of the chemical and magnetic depth profiles of films with nanometer resolution~\cite{MRFitzsimmons04,MRFitzsimmons05,LGParratt54}. The specular reflectivity, $R$, of the sample is measured as a function of wave vector transfer $Q = 4\pi \sin\theta/\lambda$, where $\theta$ is angle of incidence and $\lambda$ is x-ray/neutron wavelength. The reflectivity is related to square of the Fourier transform of the scattering length density (SLD) depth profile $\rho(z)$ (normal to the film surface or along the $z$-direction)~\cite{MRFitzsimmons05} averaged over a region typically microns in size determined by the coherence of the x-ray or neutron beams. For XRR, $\rho_x (z)$ is proportional to electron density~\cite{MRFitzsimmons05}, whereas for PNR, $\rho(z)$ consists of nuclear and magnetic SLDs such that
$\rho^± (z)=\rho_n (z) \pm  CM(z)$, where $C = 2.911\times 10^{-9}$ \AA$^{-2}$ m/kA, and $M(z)$ is the magnetization (in kA/m) depth profile. The superscript $+(-)$ sign denotes neutron beam polarization along (opposite to) the applied field.  $\rho_n (z)$ and $M(z)$ can be inferred from the experimental reflectivity data for neutron with spin parallel ($R^+$) and antiparallel ($R^-$) to magnetic field. 
The difference between $R^{+}(Q)$ and $R^{-}(Q)$ divided by the sum, called the spin asymmetry [$(R^{+}(Q)-R^{-}(Q))/(R^{+}(Q)+R^{-}(Q))$], can be a very sensitive measure of small $M$.
Figure~\ref{fig:PNR} shows the PNR data of a [(LSMO)$_6$/(BFO)$_5$]$_8$ superlattice normalized to the Fresnel reflectivity $RF=(16\pi^2)/Q^4$~\cite{MRFitzsimmons05}, where PNR measurements were performed at different temperatures while warming the sample in a field of 1 kOe after cooling it in a field of 1 kOe (FC) from 300 K.

At 300 K, the $R^+$ and $R^-$ (see Fig.~\ref{fig:PNR}(a)) are the same, indicating no net magnetization of the sample at this temperature. Fig.~\ref{fig:PNR}(b) shows the $R^{\pm}(Q)$ data obtained at 10 K for FC condition, where the calculated $R^{\pm}(Q)$ is shown by the solid curves (black and green). In comparison with the data at 300 K, it is clear that $R^+$ and $R^-$ are well separated, which indicates the ferromagnetic (FM) nature throughout the superlattice at 10 K.  The layered structure obtained from XRR (Fig. S2 in Ref.~\onlinecite{SupplMater}) was used to fit PNR data at 300 K. The corresponding nuclear SLD (or NSLD) profile for PNR data at 300 K is shown in Fig.~\ref{fig:depth_profile}(a). The NSLD profile was then fixed and the magnetization profile $M (z)$ was optimized (by minimizing the $\chi^2$ measure of error~\cite{WHPress92}) using the PNR data taken at 10 K. The solid curves in Fig.~\ref{fig:PNR} are the reflectivity that is calculated using the dynamical formalism of Parratt~\cite{LGParratt54} from the NSLD and $M(z)$ profiles shown in Fig.~\ref{fig:depth_profile}(a). In order to achieve acceptably small values of $\chi^2$, we found that the nuclear roughness of the BFO/LSMO was larger than that of the LSMO/BFO interface. Previously, asymmetries of roughness have been attributed to one oxide being polar (which terminates with a rough surface) and the other being non-polar (which terminates with a smooth surface)~\cite{NNakagawa06,TTFister08}. Interestingly, we were able to adequately represent the spin dependence of the neutron reflectivity with interfaces that have the same magnetic roughness. Notwithstanding the difference of nuclear and magnetic interface roughness, the important result is that the magnetization of the BFO layer averaged over its lateral dimensions is not zero and extends a few nanometers from the interface into the whole BFO layer (see Fig.~\ref{fig:depth_profile}(b)) at low temperatures.

%It is noted that we have also assumed different models of magnetization for BFO layer for fitting PNR data at 10 K. 
A comparison of the spin asymmetry data (PNR data at 10 K) and corresponding fit assuming three different models: (i) zero magnetization, (ii) negative (antiferromagnetic) magnetization and (iii) positive (ferromagnetic) magnetization with respect to LSMO layer, for whole BFO layer is shown in Fig.~\ref{fig:PNR}(c)-(e), respectively. The spin asymmetry fit for these models (i), (ii) and (iii) gave a reduced $\chi^2$~\cite{WHPress92} of 2.0, 2.6 and 1.3 respectively.  It is evident from Fig.~\ref{fig:PNR}(c)-(e) that the model (iii) with ferromagnetic magnetization for the whole BFO layer fits the spin asymmetry (or PNR) data well throughout Q range as compared to other two cases.  Therefore, we conclude that there is a ferromagnetic moment induced in the BFO layer.

We obtained a maximum magnetization value of $401 \pm 20$ kA/m and a minimum of $75 \pm 25$ kA/m at 10 K ($310 \pm 25$ kA/m and $30 \pm 12$ kA/m at 130 K) for the LSMO and BFO layers, respectively. The average magnetization of the whole sample was $235 \pm 25$ kA/m at 10 K ($155 \pm 20$ kA/m at 130 K), which is in agreement with a value of 210 kA/m (116 kA/m at 130 K) obtained from SQUID magnetometry as shown in Figs. S4 and S5 in Ref.~\onlinecite{SupplMater}.

\begin{figure}[t!]
\centering\includegraphics[%scale=0.28
width=1.0\linewidth,clip]{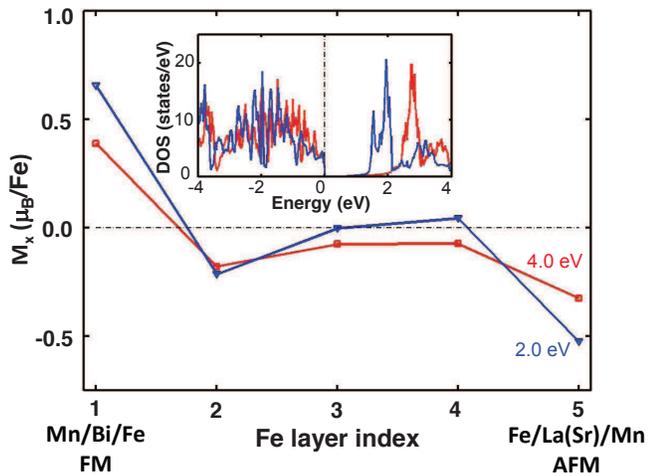}
\caption{(Color online) 
Induced ferromagnetism on Fe sites for a (LSMO)$_6$/(BFO)$_5$ system obtained from the {\it ab initio} calculations with the value of Hubbard repulsion $U = 4$ eV (red lines) and $U = 2$ eV (blue lines). The layer index 1 denotes the Fe layer interfaced with LSMO through Bi atomic layer (with the interface symbolized by Mn/Bi/Fe) while the layer index 5 denotes the Fe layer interfaced with LSMO through La(Sr) atomic layer (with the interface symbolized by Fe/La(Sr)/Mn). The inset shows the density of states for a bulk BFO with corresponding values of $U$.
}
\label{fig:lda}
\end{figure}

{\em First-principles simulations.~} To understand the origin of the magnetic interface of [(LSMO)$_6$/(BFO)$_5$]$_8$, we have performed {\em ab initio} calculations based on density functional theory using the plane-wave basis set and the projector-augmented-wave method~\cite{PEBlochl94} as implemented in the Vienna simulation package (VASP) code~\cite{GKresse96}. Calculations were performed within the local spin-density approximation plus on-site Hubbard repulsion (LSDA + $U$) on $d$-orbitals of Mn and Fe. We chose a fixed value of $U_{eff} = 4$ eV on Mn 3$d$ orbitals while varying $U_{eff}$ for Fe 3$d$ orbitals. A 500 eV energy cut-off was used to ensure the convergence of the total energy to 0.01 meV. The Brillouin zone was sampled through a mesh of $11 \times 11 \times 1$ $k$-points. The magnetization of ferromagnetic LSMO and that of the staggered magnetization of the G-type antiferromagnetic BFO were initialized to be parallel and perpendicular to  the interfacial plane. The self-consistency iteration led to a noticeable in-plane ferromagnetic moment  in BFO near the interface. Our calculations indicate that the magnitude of the induced ferromagnetic moment on Fe sites is sensitive to the size of the bandgap of BFO (shown in Fig.~\ref{fig:lda}), which is 2.6 eV in the bulk~\cite{TKanai03}. The latter, as exemplified by the density of states (shown in the inset of Fig.~\ref{fig:lda}) and depends upon the Hubbard $U$. The magnetization induced in BFO depends inversely upon the band gap. Quantitatively, the obtained interfacial ferromagnetic moment $\sim 0.3$ $\mu_{B}$/Fe ($M \sim 50$ kA/m) is much larger than the canted moment 0.03 $\mu_{B}$/Fe ($M \sim 5$ kA/m) in bulk BFO. We further show that the exchange coupling between Fe and Mn moments (along the stacking direction) across the interface is ferromagnetic when they are separated by Bi atomic layers, but the interface exchange is antiferromagnetic when they are separated by La(Sr) layers (as detailed in Fig.~\ref{fig:lda}). Ferromagnetic and antiferromagnetic exchange coupling across Bi and La(Sr) layers, respectively, explains the origin of the observed parallel and antiparallel alignment of magnetization across these layers as observed with XMCD in BFO/LSMO bilayers~\cite{PYu10} and our [(LSMO)$_6$/(BFO)$_5$]$_8$ superlattice.
 %(Fig. S3 in Ref.~\onlinecite{SupplMater}). 
We note that there is also a small magnetization canting effect in the LSMO layers, but its effect is negligible compared to that in the BFO layers. 

\begin{figure}[t]
\centering\includegraphics[%scale=0.28
width=1.0\linewidth,clip]{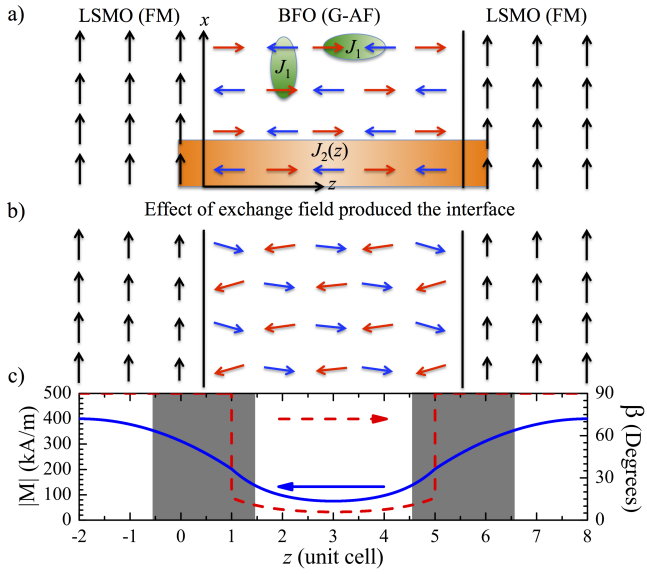}
\caption{(Color online) Field effect on BFO produced by LSMO. (a) The idealized spin configuration for the LSMO/BFO heterostructure; (b) The canted spin configuration produced by the interaction exchange field for BFO, where the general loss of magnetization in LSMO is illustrated by a change in the net moment. It should be noted that the loss magnetization could also be due to a canting of the spins in LSMO; and (c) The simulated absolute magnetization (solid-blue line) and turn angle (dashed-red line) as a function of unit cell through the LSMO/BFO heterostructure. Here, the net magnetization is assumed to be measured parallel to the interface. The gray areas show the estimated region of the interlayer mixing, which is in the range of $\sim 2$ u.c.
}
\label{fig:ex_cpl}
\end{figure}

{\em Effective exchange-coupling modeling.~} %Given both the experimental and theoretical observations of interfacial magnetization in our superlattices, where interlayer mixing, chemical asymmetry, and complex orbital interactions can potentially happen simultaneously, 
To further investigate the interfacial magnetism, we have examine
%further developed 
a classical mean-field model to investigate the exchange coupling between LSMO and BFO across the interface. Over the past few years, different theoretical studies have examined the orbital reconstruction and magnetoelastic charge transfer at the interface~\cite{MJCalderon11,SOkamoto10}. 
%Calderon et al.~\cite{MJCalderon11} proposed that magnetic moments in LSMO are orthogonal to those in the BFO layer due to the presence of orbital reconstruction at the interface~\cite{PYu10,SOkamoto10}. 
Through charge ordering and the magnetoelastic effect, the magnetic moments within the first unit cell of BFO are canted. 
%Given the interlayer mixing that is present in this system (about 1 -- 2 u.c., which is similar to the interface roughness),  
Our model allows one to consider the interlayer mixing and orbital reconstruction as an exchange field that will produce a similar canting of the local Fe moments at the interface. 
This %is particularly useful since it will produce an estimate for exchange interactions at the interface with respect to bulk BFO and 
will provide an overall energy scale for the interactions involved.
%The constructed classical exchange model is used to examine the net magnetization and canting angles of the BFO moments, where the pathways between the layers are approximated using an exchange field. 
%This provides a mean-field method to incorporate the necessary interlayer mixing and is mediated microscopically through orbital reconstruction between the Mn and Fe atoms. 
Assuming a similar interface on both sides of the BFO layer, the classical energy for the BFO layer, given as
\begin{equation}
E(\beta)= -J_1 S_1^2 \cos(2\beta)+J_2 (z) S_1 S_2  \sin\beta\;,
\end{equation}
where $J_1$ is the exchange interaction within BFO layer, $J_2 (z)$ is the exchange interaction between the moments in BFO and LSMO, which dependent on the distance $z$ measured from the interface, and $\beta$ is the angle for the BFO moments with $0^\circ$ indicating moments that are perpendicular to the interface %and $90^\circ$ indicating moments that are parallel to the interface 
(shown in Fig.~\ref{fig:ex_cpl}(a)). Here, the BFO layer is considered as an out-of-plane G-type antiferromagnet (AFM) with $S_1 = 5/2$~\cite{MMatsuda12}, while the LSMO layer is an in-plane FM with $S_2 = 2$~\cite{YSDu06}. Although these exchange parameters are not known, the strength of these parameters is, from the electronic structure point of view~\cite{JXZhu10}, dependent on the Hubbard $U$. 
To study the magnetic exchange proximity effect, we model the spatial dependence of $J_2$ as 
\begin{equation}
J_2 (z)=\frac{J_2^{(0)} d^2}{d^2+z^2 }+\frac{J_2^{(0)} d^2}{d^2+(z-L)^2 }\;,
\end{equation}
for $0 < z < 6$,  where $d$ is the unit cell size,  $L-1$ is the length of the BFO layer.  This denotes the exchange field into the BFO layer reduces as $1/z^2$.
Through an energy minimization with respect to $\beta$, the exchange field dependent $\beta$ and the associated magnetization is given by
\begin{equation}
\sin\beta= -J_2 (z)S_2/4J_1S_1 =V M(z)/g\mu_B S_1 \;,
\end{equation}
which provides the absolute magnetization $M(z)$ for at distance r from the interface. Here $g$ is the electron gyromagnetic ratio, $\mu_B$ is the Bohr magneton, and $V$ is the volume of the BFO unit cell and has the value of 61.91 \AA$^3$ for a thin film structure~\cite{JWang03}.  With $z=3d$,  $L=6d$, $g=2$, and using  the minimal  magnetization value of $75$ kA/m in the middle of BFO layer at $T=10$ K from the experiment,  we are able to determine that the exchange parameter ratio $J_2^{(0)}/J_1 = 2.5$ is needed in order for the exchange-field to produce the observed canting effect to the nearest-neighbor spins in BFO (as shown in Fig.~\ref{fig:ex_cpl}(b)).  Therefore, if we assume an exchange interaction similar to the bulk BFO ($J_1 = 4.0$ meV)~\cite{JJeong12}, then this estimates the upper limit to the exchange at the interface to be $J_{2}^{(0)} = 10$ meV. The estimation of $J_2^{(0)}$ provides an energy scale for net exchange across the interface.
In our modeling, we are able to estimate the maximal magnetization induced from the LSMO near the interface is about 202 kA/m. This value is in reasonable agreement with the experimentally observed maximal magnetization value.
%By adjusting the thickness of the BFO layer to match the experimental minimum and the interfacial magnetization, the BFO thickness was determined to be $\sim 5$ unit cells, which is consistent with the x-ray and neutron reflectometry measurements.
Figure~\ref{fig:ex_cpl}(c) shows the calculated absolute magnetization and Fe canting angle $\beta$ (the net change in angle from Figs.~\ref{fig:ex_cpl}(a-b) as a function of $z/d$ through the LSMO/BFO heterostructure.  As it is shown, the spatial dependence of magnetization is in good agreement with experimental data.

%However, PNR measurements show that the magnetization of LSMO is less than that observed in the bulk. This presents the possibility of canting of the LSMO moments or a general loss of magnetization due to an averaging effect produced by interlayer mixing\cite{LSMO-M}.  

%Since this model does not help in the determination of the overall loss of magnetization or canting in LSMO, the LSMO curvature was assumed to have similar $z^2$, that is,  $M_{FM} = M_{max} (1+\xi z^2/d^2)$ for $-3 < z < 1$ and $5 < z <9$. Here $\xi$ is a dimensionless scaling factor.  We found that 
%$\xi  = 0.03$ producing a similar curvature as the BFO magnetization, and maintaining 
%a continuous interface magnetization (Fig.~\ref{fig:ex_cpl}(c)) can be maintained with $\xi  = 0.03$. 

{\em Conclusion.}   
In summary, we have experimentally observed a significantly induced low temperature magnetization extending into several unit cells of BFO in LSMO/BFO superlattices.  Remarkable agreement between our experimental results and theory has been achieved using density functional theory and an exchange-field model through the interface of LSMO and BFO. Our {\em ab initio} calculations at low temperatures reveal the sensitive nature of induced magnetization on Fe sites at low temperatures on the band gap of BFO but also to the whether Bi or La(Sr) atomic layers are present across the LSMO/BFO interface, since this influences the exchange coupling between the Fe and Mn moments (ferromagnetic or antiferromagnetic). In this way, the relative orientations of the magnetizations of the two layers across the interface observed previously by XMCD for LSMO/BFO bilayers can be explained~\cite{PYu10}, as well as for our [(LSMO)$_6$/(BFO)$_5$]$_8$ superlattices. Since our calculations have also predicted that exchange coupling between the Fe and Mn moments can also be ferromagnetic, it will be very interesting in future to control the interface so that we can experimentally observe such an effect. Furthermore, our classical mean-field model is consistent with the understanding of orbital reconstruction at the interface and postulates a basic methodology for the determination of interaction energy scales. Overall, our work may provide the framework to address the key challenges of understanding emergent behaviors at oxide interfaces. Further measurements on the thickness dependence of magnetization will allow the picture of magnetic exchange through the interface to be clarified.
%In conclusion, by using polarized neutron reflectometry, we observed an induced magnetization of 75 ± 25 kA/m extending several unit cells into the BFO layer at 10 K. Including a classical exchange field between the LSMO and BFO layers in our density functional theory, we find the size of the BFO band gap to play an important role in determining the magnitude and length scale of the induced magnetization depends.  

{\em Acknowledgments.} 
This work was supported by the LANL/LDRD program and the Center for Integrated Nanotechnologies (CINT) at Los Alamos National Laboratory. This work has benefited from the use of the Lujan Neutron Scattering Center, which is funded by the Department of EnergyÕs Office of Basic Energy Sciences.  P.L. acknowledged the support from Sandia National Laboratories. J.L.M-D. acknowledges the ERC Advanced Investigator Grant NOVOX ERC-2009-adG247276, and EPSRC.  Authors S.S., J.T.H., and J.X. have equal contributions to the work.

\end{document}

\title{Supplemental Material for Induced magnetization in La$_{0.7}$Sr$_{0.3}$MnO$_3$/BiFeO$_3$ superlattices}

\author{S. Singh}
\affiliation{Los Alamos National Laboratory,
Los Alamos, New Mexico 87545, USA}
\affiliation{Solid State Physics Division, Bhabha Atomic Research Center, Mumbai 400085, India}

\author{J. T. Haraldsen}
\affiliation{Los Alamos National Laboratory,
Los Alamos, New Mexico 87545, USA}
\affiliation{Department of Physics and Astronomy, James Madison University, Harrisonburg, Virginia 22807, USA}

\author{J. Xiong}
\affiliation{Los Alamos National Laboratory,
Los Alamos, New Mexico 87545, USA}
\affiliation{State Key Lab of Electronic Thin Films and Integrated Devices, University of Electronic Science and Technology of China, Chengdu 610051, China} 

\author{E. M. Choi}
\affiliation{Department of Materials Science and Metallurgy, University of Cambridge, Cambridge CB2 3QZ, UK}

\author{P. Lu}
\affiliation{Sandia National Laboratory, Albuquerque, New Mexico 87185, USA}

\author{ D. Yi}
\affiliation{Department of Materials Science and Engineering, University of California at Berkeley, Berkeley, California 94720, USA }

\author{X.-D. Wen}
\affiliation{Los Alamos National Laboratory,
Los Alamos, New Mexico 87545, USA}

\author{J. Liu}
\affiliation{Department of Materials Science and Engineering, University of California at Berkeley, Berkeley, California 94720, USA }

\author{H. Wang}
\affiliation{Department of Electrical and Computer Engineering, Texas A\&M University, College Station, Texas 77843, USA}

\author{Z. Bi}
\affiliation{Los Alamos National Laboratory,
Los Alamos, New Mexico 87545, USA}

\author{P. Yu}
\affiliation{Department of Materials Science and Engineering, University of California at Berkeley, Berkeley, California 94720, USA }

\author{M. R. Fitzsimmons}
\affiliation{Los Alamos National Laboratory,
Los Alamos, New Mexico 87545, USA}

\author{J. L. MacManus-Driscoll}
\affiliation{Department of Materials Science and Metallurgy, University of Cambridge, Cambridge CB2 3QZ, UK}

\author{R. Ramesh}
\affiliation{Department of Materials Science and Engineering, University of California at Berkeley, Berkeley, California 94720, USA }

\author{A. V. Balatsky}
\affiliation{Los Alamos National Laboratory,
Los Alamos, New Mexico 87545, USA}
%\affiliation{Center for Integrated Nanotechnologies, Los Alamos National Laboratory, Los Alamos, New Mexico 87545, USA}

\author{Jian-Xin Zhu}
\affiliation{Los Alamos National Laboratory,
Los Alamos, New Mexico 87545, USA}

\author{Q. X. Jia}
\affiliation{Los Alamos National Laboratory,
Los Alamos, New Mexico 87545, USA}

%\pacs{71.27.+a, 74.55.+v, 75.20.Hr, 75.30.Mb}
%71.27.+a	Strongly correlated electron systems; heavy fermions
%74.55.+v	Tunneling phenomena: single particle tunneling and STM
%75.20.Hr	Local moment in compounds and alloys; Kondo effect, valence fluctuations, heavy fermions
%75.30.Mb	Valence fluctuation, Kondo lattice, and heavy-fermion phenomena
\maketitle

%\onecolumngrid
%\twocolumngrid
%\section*{Supplemental Material for EPAPS}

\section{Sample Growth}

%\begin{equation}
%x=y
%\end{equation}
%
%\begin{subequations}
%\begin{eqnarray}
%A&=&B\;, \\
%C&=& D\;.
%\end{eqnarray}
%\end{subequations}

Pulsed laser (KrF) deposition was used to grow superlattices [(LSMO)$_n$/(BFO)$_m$]$_N$ on (001) SrTiO$_3$ (STO) substrates, where n = 2, 4, 6, 8, m = 2, 4, 5, 8, and N = 6, 8, 12, 24. The deposition rate was controlled through appropriate focus of laser beam on the target, where the nominal growth rate was $\sim 0.01$ \AA/pulse. The substrate temperature during film growth was initially optimized and was maintained at 750 $^\circ$C as calibrated by the pyrometer. The oxygen pressure during deposition was 200 mTorr. The samples were cooled to room temperature in pure oxygen (300 Torr) by turning off the power supply to the heater without further thermal treatment.

%\begin{figure}[h!]
%\centering
%\subfigure[\ Real and imaginary.]{
%\includegraphics[scale=0.5]{reim}}
%\subfigure[\ Amplitude and phase.]{
%\includegraphics[scale=0.5]{phase}}
%\caption{\label{fig:qm/complexfunctions} Two representations of complex
%wave functions.}
%\end{figure}

%\onecolumngrid
%\twocolumngrid

\section{Sample Characterization}

\begin{figure}[b]
\centering\includegraphics[%scale=0.28
width=1.0\linewidth,clip]{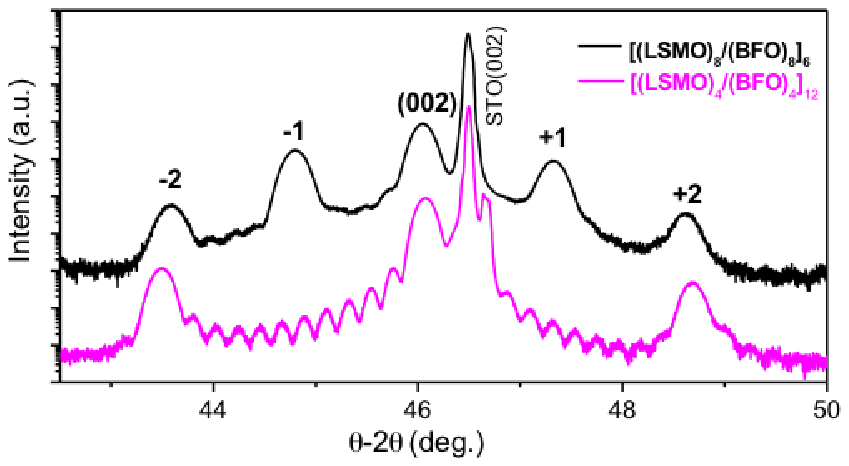}
\caption{(Color online) 
x-ray $\theta$-$2\theta$ scans of [(LSMO)$_8$/(BFO)$_8$]$_6$ and [(LSMO)$_4$/(BFO)$_4$]$_{12}$ superlattices.
}
\label{Sfig:XRS}
\end{figure}

\begin{figure}[b]
\centering\includegraphics[%scale=0.28
width=1.0\linewidth,clip]{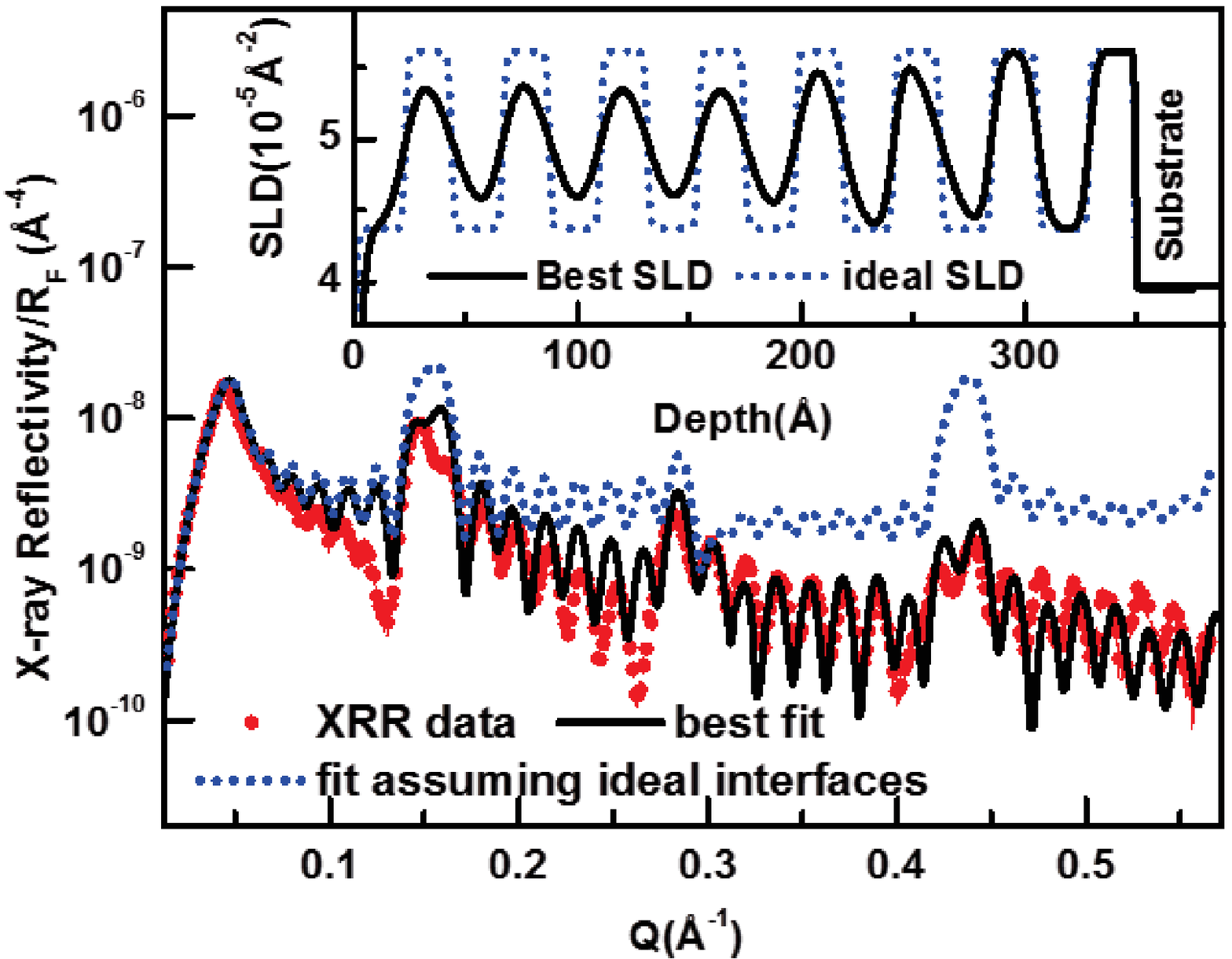}
\caption{(Color online) 
XRR pattern of a superlattice [(LSMO)$_6$/(BFO)$_5$]$_8$.
}
\label{Sfig:XRR}
\end{figure}

The crystalline quality of the superlattice and its epitaxial registry with the SrTiO$_3$ substrate was evaluated by Rutherford Backscattering Spectroscopy in the channeling mode using 2 MeV He+ ions. Both x-ray normal $\theta$-$2\theta$ scans and x-ray reflectivity (XRR) were used to characterize the superlattice. Figure~\ref{Sfig:XRS} shows the typical x-ray $\theta$-$2\theta$ scans of [(LSMO)$_8$/(BFO)$_8$]$_6$ and [(LSMO)$_4$/(BFO)$_4$]$_{12}$ superlattices. Both thickness fringes and superlattice peaks are clearly visible. The diffraction (shown in Fig.~\ref{Sfig:XRR}) exhibits a periodic sequence of the satellite maxima. These results are evidence for a high degree of perfection of atomic structure along the growth direction. Figure S2 shows a typical XRR spectrum of a superlattice of [(LSMO)$_6$/(BFO)$_5$]$_8$, where the reflectivity has been normalized to the Fresnel reflectivity ($RF =  16\pi^2/Q^4$)~\cite{Fitzsimmons05}.  By fitting the XRR data, we obtained the thicknesses for each layer of BFO and LSMO of $20 \pm 1\;{\text  \AA}$  and $23.0 \pm 1$ \AA, respectively, which is consistent with the expected thicknesses of the BFO (5 u.c., $c = 3.962$ \AA) and LSMO (6 u.c., $c = 3.870$ \AA). Inset in the Fig. S2 shows the layer structure (electron scattering length density (ESLD) depth profile) which gave best fit to XRR data. The layer structure was then used to fit neutron reflectivity data.

\begin{figure}[h]
\centering\includegraphics[%scale=0.28
width=1.0\linewidth,clip]{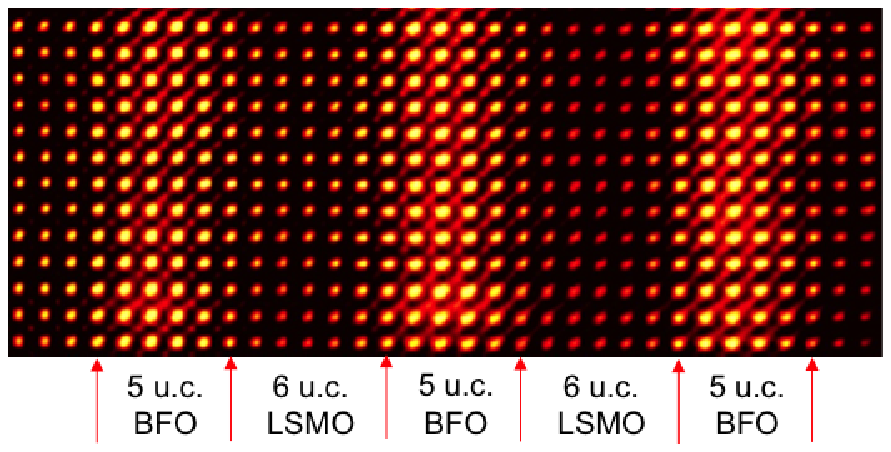}
\caption{(Color online) 
High angle annular dark field (HAADF) image showing the interfaces between BFO and LSMO or LSMO and BFO. The image was acquired in the [100] zone axis with an electron probe of convergence angle of 18.5 mrad and size of less than 1.5 \AA and an annular detector of collection range of 60 - 160 mrad. u.c. = unit cell.
}
\label{Sfig:HAADF}
\end{figure}

\begin{figure}[ht]
\centering\includegraphics[%scale=0.28
width=1.0\linewidth,clip]{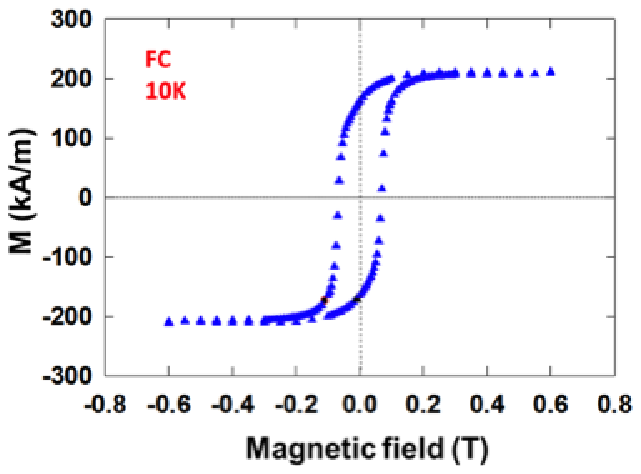}
\caption{(Color online) 
Magnetization vs. magnetic field hysteresis at 10 of the superlattice [(LSMO)$_6$/(BFO)$_5$]$_8$.
}
\label{Sfig:MvB}
\end{figure}

\begin{figure}[ht]
\centering\includegraphics[%scale=0.28
width=1.0\linewidth,clip]{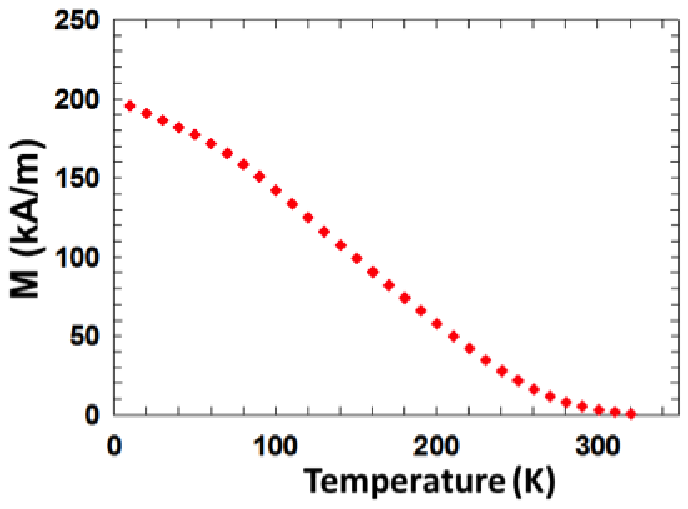}
\caption{(Color online) 
XRR pattern of a superlattice [(LSMO)$_6$/(BFO)$_5$]$_8$.
}
\label{Sfig:MvT}
\end{figure}

%\twocolumngrid

An FEI Titan TM G2 80-200 STEM with a Cs probe corrector and ChemiSTEM TM technology (X-FEGTM and SuperX TM EDS with 4 windowless silicon drift detectors), operated at 200 kV, was used to evaluate sharpness across the interface.  High angle annular dark field (HAADF) image was acquired in the [100] zone axis with an electron probe of convergence angle of 18.5 mrad, a size of less than 1.5 \AA, and an annular detector of collection range of 60 - 160 mrad. The TEM sample was prepared by focused ion beam (FIB), which initially produces the thick TEM specimen of thickness greater than 1 $\mu$m, followed by Ar ion milling at 3 kV to reduce the specimen thickness. The Ar ion-milling at 3 kV was used to remove the FIB damages to the specimen. Figure~\ref{Sfig:HAADF} shows the HAADF Z-contrast image of the superlattice [(LSMO)$_6$/(BFO)$_5$]$_8$. It can be seen that the superlattice has well-defined interfaces between the LSMO and the BFO components. The thickness and the periodicity obtained from the TEM agree with the x-ray diffraction analysis.

The macroscopic magnetization hysteresis of the superlattice was measured using a superconducting quantum interference device (SQUID) magnetometer. Figures~\ref{Sfig:MvB} and~\ref{Sfig:MvT} show the magnetization vs. magnetic field hysteresis at 10 K and magnetization vs. temperature of the superlattice [(LSMO)$_6$/(BFO)$_5$]$_8$ respectively, where field cooled (FC) measurement was done by initially cooling the sample in a field of 1 kOe from 300 K, and then measuring the magnetization while warming the sample at the same field. The magnetization of the sample was 210 kA/m at 10 K (116 kA/m at 130 K), which is in agreement with a value of $235 \pm 25$ kA/m at 10 K ($155 \pm  20$ kA/m at 130 K) measured from our PNR measurements.